\pdfoutput=1
\documentclass[prl,twocolumn,showpacs,aps]{revtex4}
\usepackage{graphicx,graphics,color,epsfig}
\usepackage{bm}
\usepackage{amsmath}
\usepackage{amssymb}
\usepackage{epstopdf}

\makeatletter

\newcommand{\Rmnum}[1]{\expandafter\@slowromancap\romannumeral #1@}
\makeatother

\usepackage{color}

\begin{document}

\title{Hidden sign-changing $s$-wave superconductivity in monolayer FeSe}

\author{Yi Gao,$^{1}$ Yan Yu,$^{1}$ Tao Zhou,$^{2}$ Huaixiang Huang,$^{3}$ and Qiang-Hua Wang $^{4,5}$}
\affiliation{$^{1}$Department of Physics, Nanjing Normal University, Nanjing, 210023, China\\
$^{2}$College of Science, Nanjing University of Aeronautics and Astronautics, Nanjing, 210016, China\\
$^{3}$Department of Physics, Shanghai University, Shanghai, 200444, China\\
$^{4}$National Laboratory of Solid State Microstructures $\&$ School of Physics, Nanjing
University, Nanjing, 210093, China\\
$^{5}$Collaborative Innovation Center of Advanced Microstructures, Nanjing 210093, China}

\begin{abstract}
Combining the recent scanning tunneling microscopy (STM) and angle-resolved photoemission spectroscopy (ARPES) measurements, we construct a tight-binding model suitable for describing the band structure of monolayer FeSe grown on SrTiO$_{3}$. Then we propose a possible pairing function, which can well describe the gap anisotropy observed by ARPES and has a hidden sign-changing characteristic. At last, as a test of this pairing function we further study the nonmagnetic impurity-induced bound states, to be verified by future STM experiments.

\end{abstract}


\maketitle

Recently, one monolayer (1ML) FeSe thin film grown on SrTiO$_{3}$ has attracted much attention due to its intriguing interfacial properties and high superconducting (SC) transition temperature T$_{c}$ \cite{xueqk1, zhouxj1, zhouxj2,fengdl1,fengdl2,fengdl3,xueqk2,shenzx1,jiajf,chenxh,hoffman}. On the one hand, its properties are drastically different from bulk FeSe. For example, compared to bulk FeSe whose T$_{c}$ is below 10K \cite{wumk}, T$_{c}$ in 1ML FeSe can exceed 50K. Meanwhile, the SC gap in 1ML FeSe can be as large as 10-20meV, in contrast to its value in bulk FeSe (below 3.5meV) \cite{matsuda}. Furthermore, the SC gap structure in bulk FeSe shows a nodal behavior \cite{xueqk3} whereas full gap opens in 1ML FeSe.

On the other hand, 1ML FeSe is quite different from most other iron-based superconductors as well. Detailed investigation of the electronic structure by angle-resolved photoemission spectroscopy (ARPES) shows that, in 1ML FeSe, there are only electron pockets around the $M$ points of the 2Fe/cell Brillouin zone (BZ), while the usual hole pockets around $\Gamma$ in most iron-based superconductors sink and are located at about 80meV below the Fermi energy, leaving no pockets around $\Gamma$ \cite{fengdl1,fengdl2,shenzx1,zhouxj1,zhouxj2}. In this case, 1ML FeSe is about $10\%$ electron doped and the Fermi momentum $k_{F}/\pi\approx0.25$ \cite{zhouxj1}.

Up to now, various theories have been constructed to account for the SC mechanism and pairing symmetry in 1ML FeSe. In most iron-based superconductors with hole pockets around $\Gamma$, it has been widely accepted that the pairing symmetry is $s_{\pm}$-wave. The pairing order parameter $\Delta_{\mathbf{k}}$ changes sign between the $\Gamma$ hole pockets and the $M$ electron pockets while it can be qualitatively described as $\Delta_{\mathbf{k}}\sim \cos k_{x}+\cos k_{y}$ (or $\cos k_{x}\cos k_{y}$ defined in the 1Fe/cell BZ), corresponding to the next-nearest-neighbor (NNN) pairing between the Fe atoms. However for 1ML FeSe, the situation is completely different. At the beginning, it was suggested that, if the SC mechanism is due to the spin fluctuation (resulted from the electron-electron correlation), then since there are no hole pockets, the pairing symmetry should be nodeless $d$-wave \cite{maier,leedh}. Later it was found that, with the electron-phonon interaction between 1ML FeSe and the SrTiO$_{3}$ substrate, the pairing symmetry may change from $d$-wave to $s$-wave and this $s$-wave symmetry can be thought of as the usual $s_{\pm}$ symmetry restricted to the exposed electron pockets \cite{wangqh}.

Experimentally, by measuring the quasiparticle interference in the presence and absence of a magnetic field, Ref. \cite{fengdl3} rules out any sign change of the pairing order parameter on the Fermi surfaces and excludes the $d$-wave pairing symmetry. The latest high-resolution ARPES found that, there are two electron pockets $\delta_{1}$ and $\delta_{2}$ around $M$. The SC gap on the outer pocket $\delta_{2}$ is slightly larger than that on the inner one $\delta_{1}$. The gap is anisotropic with its maxima located along the $\Gamma-M$ line and minima located along the $X-M$ line \cite{shenzx2}. The authors show that the usual $s_{\pm}$ symmetry (even if restricted to the exposed electron pockets) is not consistent with their data since the gap minima and maxima are located at wrong positions. They further inferred that mixing of different gap functions with the same symmetry may explain the observed gap anisotropy.

Combining the above mentioned scanning tunneling microscopy (STM) and ARPES measurements \cite{fengdl3,shenzx2}, in this work, first we construct a tight-binding model suitable for describing the band structure of 1ML FeSe, then we propose a possible pairing function, which can well describe the gap anisotropy observed by ARPES. At last, based on this pairing function we further study the nonmagnetic impurity-induced in-gap bound states as a test of our pairing function, which can be verified by future STM experiments.

In most iron-based superconductors, the Fe atoms form a two-dimensional lattice, with the Se/As atoms located below and above the Fe plane at exactly the same distance. Therefore the glide mirror symmetry [$z\leftrightarrow-z$ reflection with respect to the Fe plane followed by a translation to nearest-neighbor (NN) Fe] is present, in this case people can take only the Fe atoms into account and work in the 1Fe/cell BZ. Then by using a folding scheme to fold the band into the 2Fe/cell BZ, people can compare the calculated band structure to that observed by ARPES.  However for 1ML FeSe grown on SrTiO$_{3}$, the glide mirror symmetry is explicitly broken since the up and down Se atoms reside in completely different environment and their distances to the Fe plane may differ. Thus the folding scheme does not work anymore and any tight-binding model must take this into account and must be built in the 2Fe/cell BZ at the first place. Besides, since we are mostly interested in the STM and ARPES experiments, which mainly probe the surface properties of materials where the glide mirror symmetry is also broken. Therefore we follow the idea proposed in Ref. \cite{zhangdg} and build a phenomenological model in the 2Fe/cell BZ to fit the basic characteristics of the ARPES-measured band structure. Previously this idea has been successfully applied to explain the vortex states observed in STM experiment \cite{gaoy,wenhh}.

\begin{figure}
\includegraphics[width=1\linewidth]{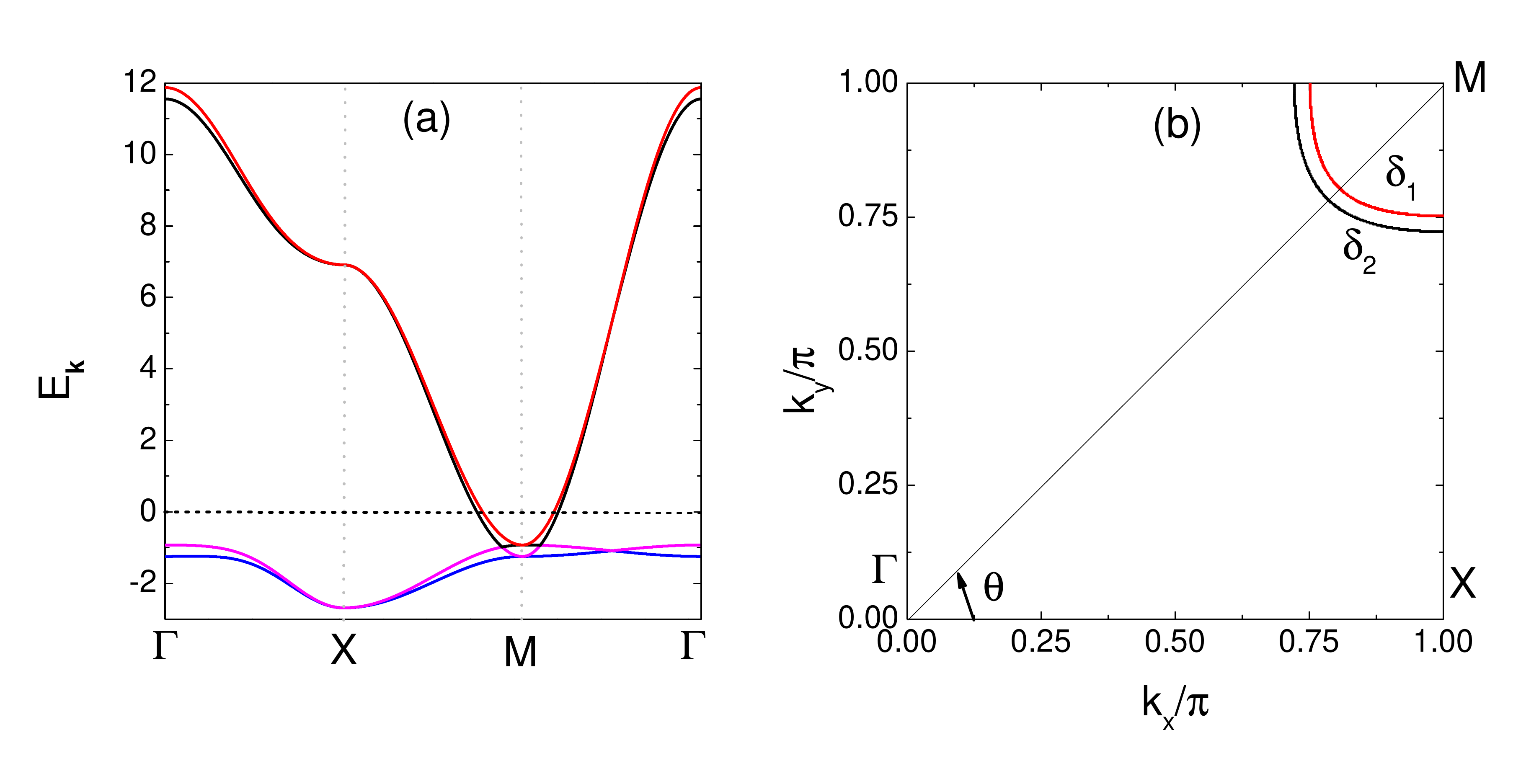}
 \caption{\label{band} (color online) (a) Calculated band structure along the high-symmetry directions in the 2Fe/cell BZ. The energy is defined with respect to the Fermi energy (the black dotted line). (b) The Fermi surfaces in the first quadrant of the 2Fe/cell BZ.}
\end{figure}

In the two-dimensional Fe lattice, since the broken of the glide mirror symmetry, each unit cell contains two inequivalent sublattices $A$ and $B$. The coordinate of the sublattice $A$ in the unit cell $(i,j)$ is $\mathbf{R}_{ij}=(i,j)$ while that for the sublattice $B$ is $\mathbf{R}_{ij}+\mathbf{d}$, with $\mathbf{d}$ being $(0.5,0.5)$. The Hamiltonian can be written as
\begin{eqnarray}
\label{h}
H&=&\sum_{\mathbf{k}}\psi_{\mathbf{k}}^{\dag}A_{\mathbf{k}}\psi_{\mathbf{k}},\nonumber\\
\psi_{\mathbf{k}}^{\dag}&=&(c_{\mathbf{k}A1\uparrow}^{\dag},c_{\mathbf{k}A2\uparrow}^{\dag},c_{\mathbf{k}B1\uparrow}^{\dag},c_{\mathbf{k}B2\uparrow}^{\dag},\nonumber\\
&&c_{-\mathbf{k}A1\downarrow},c_{-\mathbf{k}A2\downarrow},c_{-\mathbf{k}B1\downarrow},c_{-\mathbf{k}B2\downarrow}),\nonumber\\
A_{\mathbf{k}}&=&\begin{pmatrix}
M_{\mathbf{k}}&\Delta_{\mathbf{k}}\\\Delta_{\mathbf{k}}^{\dag}&-M_{-\mathbf{k}}^{T}
\end{pmatrix},\nonumber\\
M_{\mathbf{k}}&=&\begin{pmatrix}
\epsilon_{A,\mathbf{k}}&\epsilon_{xy,\mathbf{k}}&\epsilon_{T,\mathbf{k}}&0\\
\epsilon_{xy,\mathbf{k}}&\epsilon_{A,\mathbf{k}}&0&\epsilon_{T,\mathbf{k}}\\
\epsilon_{T,\mathbf{k}}&0&\epsilon_{B,\mathbf{k}}&\epsilon_{xy,\mathbf{k}}\\
0&\epsilon_{T,\mathbf{k}}&\epsilon_{xy,\mathbf{k}}&\epsilon_{B,\mathbf{k}}
\end{pmatrix},
\end{eqnarray}
where $c_{\mathbf{k}A1\uparrow}^{\dag}/c_{\mathbf{k}A2\uparrow}^{\dag}$ creates a spin up electron with momentum $\mathbf{k}$ and on the $d_{xz}/d_{yz}$ orbital of the sublattice $A$. $\epsilon_{A,\mathbf{k}}=-2(t_{2}\cos k_{x}+t_{3}\cos k_{y})-\mu$, $\epsilon_{B,\mathbf{k}}=-2(t_{2}\cos k_{y}+t_{3}\cos k_{x})-\mu$, $\epsilon_{xy,\mathbf{k}}=-2t_{4}(\cos k_{x}+\cos k_{x})$ and $\epsilon_{T,\mathbf{k}}=-4t_{1}\cos(k_{x}/2)\cos(k_{y}/2)$. The broken of the glide mirror symmetry is manifested as $t_{2}\neq t_{3}$ since they are the NNN hoppings mediated by the up and down Se. In addition, $M_{\mathbf{k}}$ and $\Delta_{\mathbf{k}}$ are the tight-binding and pairing parts of the system, respectively. Throughout this work, the momentum $\mathbf{k}$ is defined in the 2Fe/cell BZ. In the following we set $t_{1-4}=1.6,0.4,-2,0.04$ and $\mu=-1.9$ to fit the band structure measured by ARPES. Under this set of parameters, the average electron occupation number is $n\approx2.1$, leading the system to be about $10\%$ electron doped. The calculated band structure and Fermi surfaces are shown in Fig. \ref{band}. As we can see, the $\Gamma$ hole pockets sink below the Fermi energy while two electron pockets $\delta_{1}$ and $\delta_{2}$ exist around $M$ with their sizes similar to the ARPES data ($k_{F}/\pi\approx0.25$). Therefore, both the electron number and the Fermi surface topology are consistent with the ARPES measurements \cite{zhouxj1,shenzx2}.

\begin{figure}
\includegraphics[width=1\linewidth]{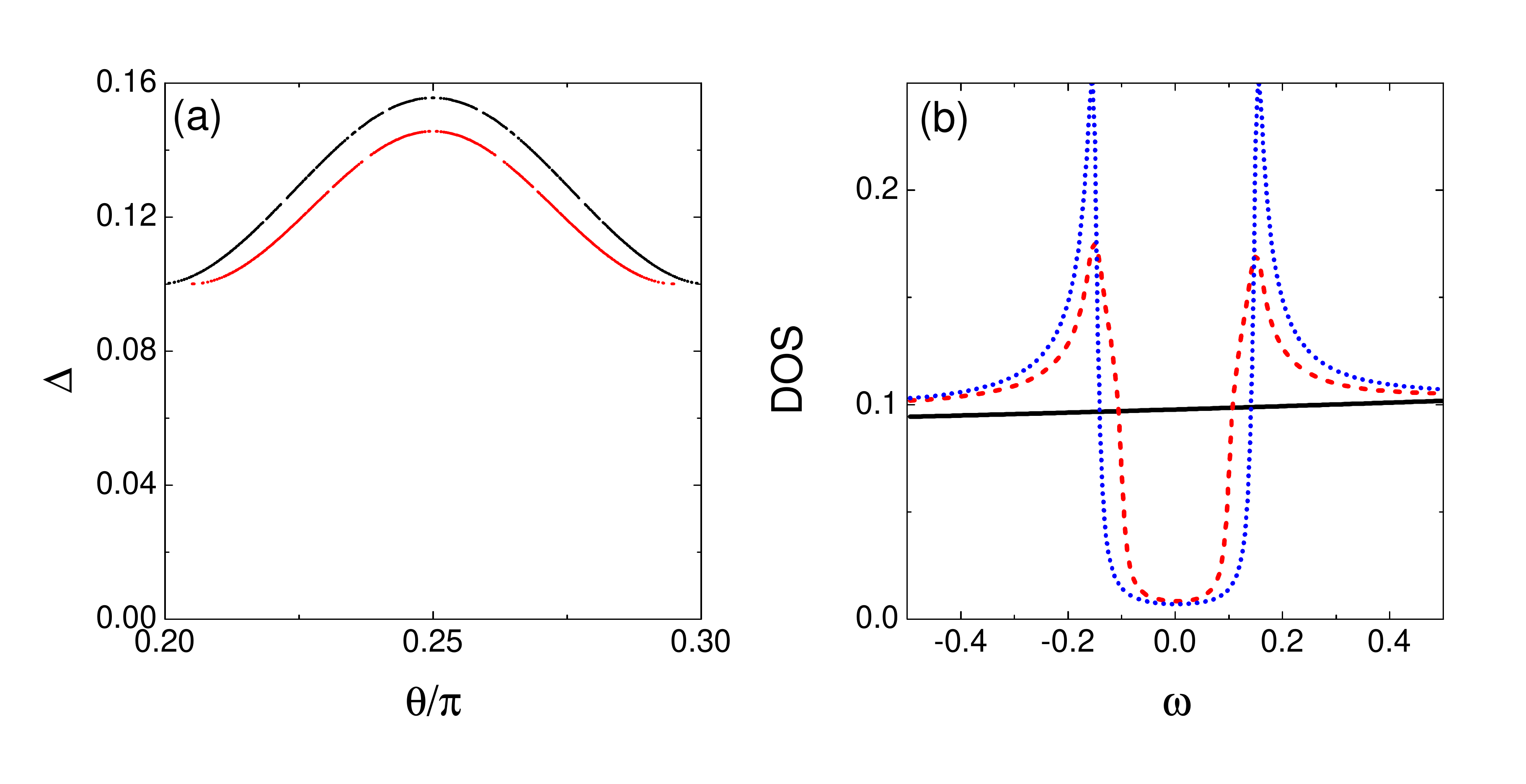}
 \caption{\label{gap&dos} (color online) (a) The magnitude of the SC gap along $\delta_{1}$ (red) and $\delta_{2}$ (black). (b) The DOS in the normal state (black solid), the $s^{*}$-wave pairing state (red dashed) and the conventional $s$-wave pairing state (blue dotted).}
\end{figure}

Then we come to the pairing function $\Delta_{\mathbf{k}}$. From Ref. \cite{fengdl3} we know that $\Delta_{\mathbf{k}}$ should have generally an $s$-wave symmetry. Ref. \cite{shenzx2} shows that the SC gap is larger on $\delta_{2}$ than it is on $\delta_{1}$ while on both $\delta_{1}$ and $\delta_{2}$, the gap maxima are located along the $\Gamma-M$ line [at $\theta=\pi/4$ where $\theta$ is defined in Fig. \ref{band}(b)] and the minima are located along the $X-M$ line. Combining all these above, we propose that $\Delta_{\mathbf{k}}$ can be written as
\begin{eqnarray}
\label{dk}
\Delta_{\mathbf{k}}&=&\begin{pmatrix}
\Delta_{0}&0&\Delta_{1\mathbf{k}}&0\\
0&\Delta_{0}&0&\Delta_{1\mathbf{k}}\\
\Delta_{1\mathbf{k}}&0&\Delta_{0}&0\\
0&\Delta_{1\mathbf{k}}&0&\Delta_{0}
\end{pmatrix},
\end{eqnarray}
where $\Delta_{0}=-0.1$ and $\Delta_{1\mathbf{k}}=0.5\cos(k_{x}/2)\cos(k_{y}/2)$. Here $\Delta_{0}$ is momentum-independent and originates from the on-site intraorbital pairing, with the pairing symmetry being conventional $s$-wave. On the other hand, $\Delta_{1\mathbf{k}}$ is due to the NN intraorbital pairing (inter-sublattice) and its symmetry is also $s$-wave. Generally speaking, if the pairing mechanism is due to the spin fluctuation, then since the $\Gamma$ hole pockets are absent, the electrons can only be scattered between the electron pockets and the low-energy effective interaction can well be described by a $J_{1}-J_{2}$ model with the NN interaction $J_{1}$ being dominant. In this case, $\Delta_{1\mathbf{k}}$ should have the form factor $\sin(k_{x}/2)\sin(k_{y}/2)$ and the symmetry should be $d$-wave \cite{leedh}. However, there are both experimental and theoretical evidences suggesting that the electron-phonon interaction between 1ML FeSe and the SrTiO$_{3}$ substrate plays a vital role in boosting the SC gap magnitude and T$_{c}$ \cite{shenzx1,leedh2}. The electron-phonon interaction may produce an effective on-site pairing interaction and this interaction may suppress the $\sin(k_{x}/2)\sin(k_{y}/2)$ component in $\Delta_{1\mathbf{k}}$ and finally change $\Delta_{1\mathbf{k}}$ into $\cos(k_{x}/2)\cos(k_{y}/2)$. Meanwhile it results in the on-site pairing $\Delta_{0}$. The transition of the NN pairing symmetry induced by the onsite pairing has also been found in other systems \cite{gaoy2}. In Fig. \ref{gap&dos}(a) we plot the magnitude of the SC gap on $\delta_{1}$ and $\delta_{2}$. We can see that the pocket $\delta_{2}$ has a slightly larger gap magnitude than $\delta_{1}$ while on both $\delta_{1}$ and $\delta_{2}$, the gap maxima are at $\theta=\pi/4$ and the minima are located along the $X-M$ line. Furthermore, the gap minima on these two pockets are equal to each other since along $X-M$, we have $\Delta_{1\mathbf{k}}=0$ ($k_{x}=\pi$ or $k_{y}=\pi$). All the characteristics of the magnitude and distribution of the SC gaps agree quite well with the ARPES measurement \cite{shenzx2}. In addition, this pairing symmetry is different from most iron-based superconductors since their $\Delta_{1\mathbf{k}}\sim(\cos k_{x}+\cos k_{y})$, which is resulted from the NNN pairing (intra-sublattice). However the gap distribution of the NNN pairing is not consistent with the ARPES data (see Fig. 4 of Ref. \cite{shenzx2} and Fig. 4 of Ref. \cite{wangqh2}).

A closer inspection of Eq. (\ref{dk}) shows that, the phase difference between $\Delta_{0}$ and $\Delta_{1\mathbf{k}}$ is $\pi$, that is, if we change $\Delta_{0}$ into $0.1$, then the gap distribution along $\delta_{1}$ and $\delta_{2}$ will not match the ARPES data. The $\pi$ phase difference leads to the following consequences. Since $\delta_{1}$ and $\delta_{2}$ are close to $M$ where $(k_{x},k_{y})=(\pi,\pi)$, on these two pockets $|\Delta_{1\mathbf{k}}|$ is tiny and $|\Delta_{1\mathbf{k}}|\ll|\Delta_{0}|$, therefore on $\delta_{1}$ and $\delta_{2}$, the sign of the SC gap follows that of $\Delta_{0}$ ($-$). However on the bands close to $\Gamma$ where $(k_{x},k_{y})=(0,0)$, we have $|\Delta_{1\mathbf{k}}|\gg|\Delta_{0}|$ and the sign of the SC gap on those bands follows the sign of $\Delta_{1\mathbf{k}}$ there ($+$). Since the bands close to $\Gamma$ are below the Fermi energy, thus we denote this pairing symmetry as a hidden sign-changing $s$-wave symmetry $s^{*}$. This pairing symmetry is fully gapped and its density of states (DOS) can be found in Fig. \ref{gap&dos}(b), which shows a $U$-shaped profile near $\omega=0$.

In the following, we study the in-gap bound states around a nonmagnetic impurity, in order to distinguish the $s^{*}$-wave symmetry from the conventional $s$-wave symmetry where this is no sign change. When a single nonmagnetic impurity (behaves as a potential scatterer) is placed at the sublattice $A$ in the unit cell $(0,0)$, the impurity Hamiltonian can be written as
\begin{eqnarray}
H_{imp}&=&V_{s}\sum_{\alpha=1}^{2}\sum_{\sigma=\uparrow,\downarrow}c_{00A\alpha\sigma}^{\dag}c_{00A\alpha\sigma}\nonumber\\
&=&\frac{V_{s}}{N}\sum_{\alpha=1}^{2}\sum_{\sigma=\uparrow,\downarrow}\sum_{\mathbf{k},\mathbf{k}^{'}}c_{\mathbf{k}A\alpha\sigma}^{\dag}c_{\mathbf{k}^{'}A\alpha\sigma},
\end{eqnarray}
where $N$ is the number of the unit cells and $V_{s}$ is the scattering strength of the nonmagnetic impurity. Following the standard $T$-matrix procedure \cite{zhujx}, we define the Green's function matrix as $g(\mathbf{k},\mathbf{k}^{'},\tau)=-\langle T_{\tau}\psi_{\mathbf{k}}(\tau)\psi_{\mathbf{k}^{'}}^{\dag}(0)\rangle$
and
\begin{eqnarray}
\label{gw}
g^{r/a}(\mathbf{k},\mathbf{k}^{'},\omega)&=&\delta_{\mathbf{k}\mathbf{k}^{'}}g_{0}^{r/a}(\mathbf{k},\omega)\nonumber\\
&+&g_{0}^{r/a}(\mathbf{k},\omega)T^{r/a}(\omega)g_{0}^{r/a}(\mathbf{k}^{'},\omega).
\end{eqnarray}
Here $r$ and $a$ refer to the retarded and advanced Green's function, respectively and
\begin{eqnarray}
\label{g0}
g_{0}^{r/a}(\mathbf{k},\omega)&=&[(\omega\pm i0^{+})I-A_{\mathbf{k}}]^{-1},\nonumber\\
T^{r/a}(\omega)&=&[I-\frac{U}{N}\sum_{\mathbf{q}}g_{0}^{r/a}(\mathbf{q},\omega)]^{-1}\frac{U}{N},
\end{eqnarray}
where $I$ is a $8\times8$ unit matrix and the nonzero elements of the matrix $U$ are $U_{11}=U_{22}=-U_{55}=-U_{66}=V_{s}$.
The experimentally measured local density of states (LDOS) at the sublattice $A$ is expressed as
\begin{eqnarray}
\label{roura}
\rho_{A}(\mathbf{R}_{ij},\omega)&=&-\frac{1}{\pi}\sum_{\alpha=1}^{2}\sum_{\sigma=\uparrow,\downarrow}{\rm Im}\langle\langle c_{ijA\alpha\sigma}|c_{ijA\alpha\sigma}^{\dag}\rangle\rangle_{\omega+i0^{+}}\nonumber\\
&=&-\frac{1}{\pi N}\sum_{\alpha=1}^{2}\sum_{\mathbf{k},\mathbf{k}^{'}}{\rm Im}\Big{\{}[g_{\alpha\alpha}^{r}(\mathbf{k},\mathbf{k}^{'},\omega)\nonumber\\
&-&g_{\alpha+4\alpha+4}^{a}(\mathbf{k},\mathbf{k}^{'},-\omega)]e^{-i(\mathbf{k}-\mathbf{k}^{'})\cdot\mathbf{R}_{ij}}\Big{\}},
\end{eqnarray}
and similar expressions can be derived for the sublattice $B$.

\begin{figure}
\includegraphics[width=1\linewidth]{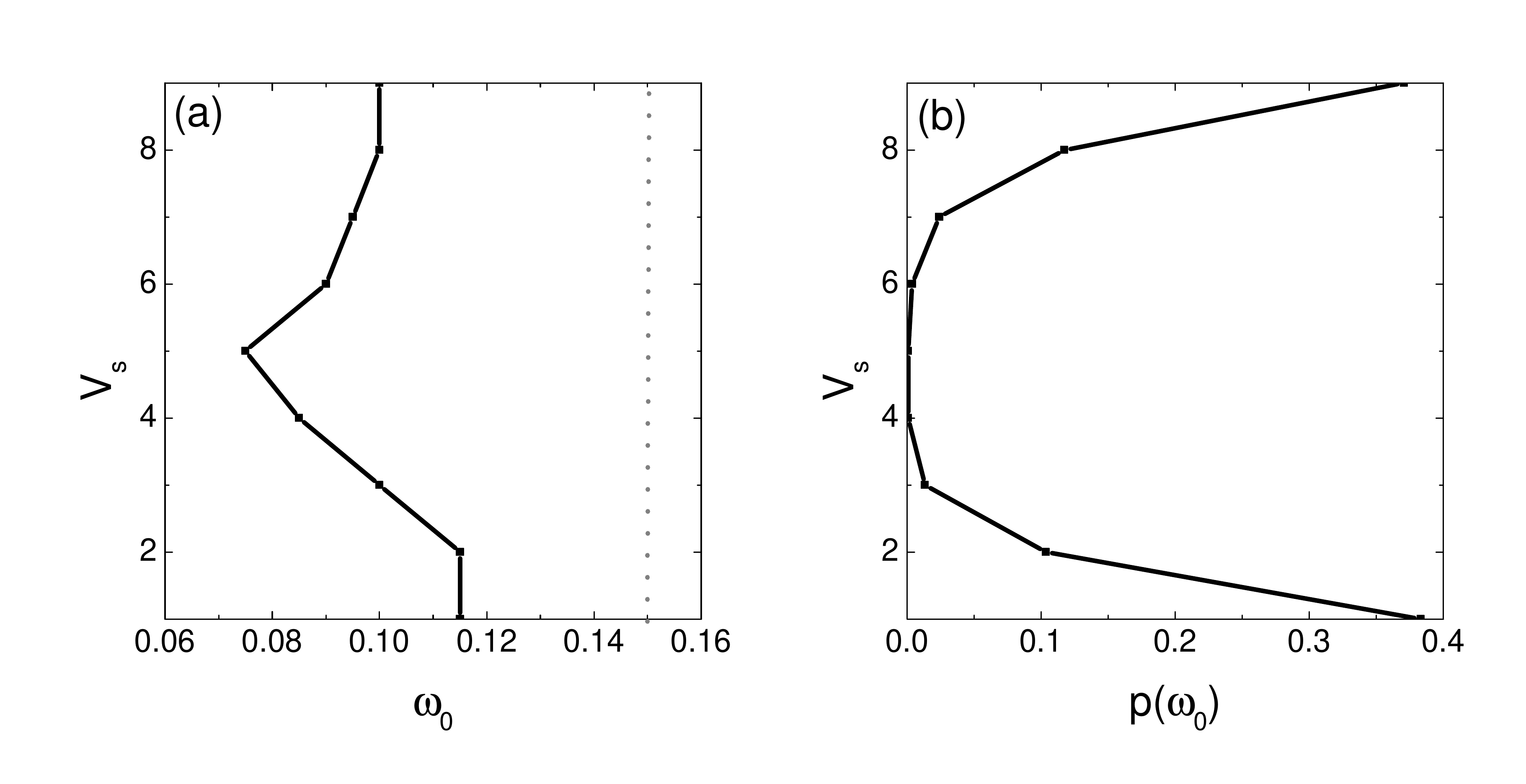}
 \caption{\label{location1} For the $s^{*}$-wave pairing symmetry. (a) $\omega_{0}$ as a function of $V_{s}$. The gray dotted line denotes the location of the SC coherence peaks. (b) $p(\omega_{0})$ as a function of $V_{s}$. See the definition of $\omega_{0}$ and $p(\omega_{0})$ in the text.}
\end{figure}

\begin{figure}
\includegraphics[width=1\linewidth]{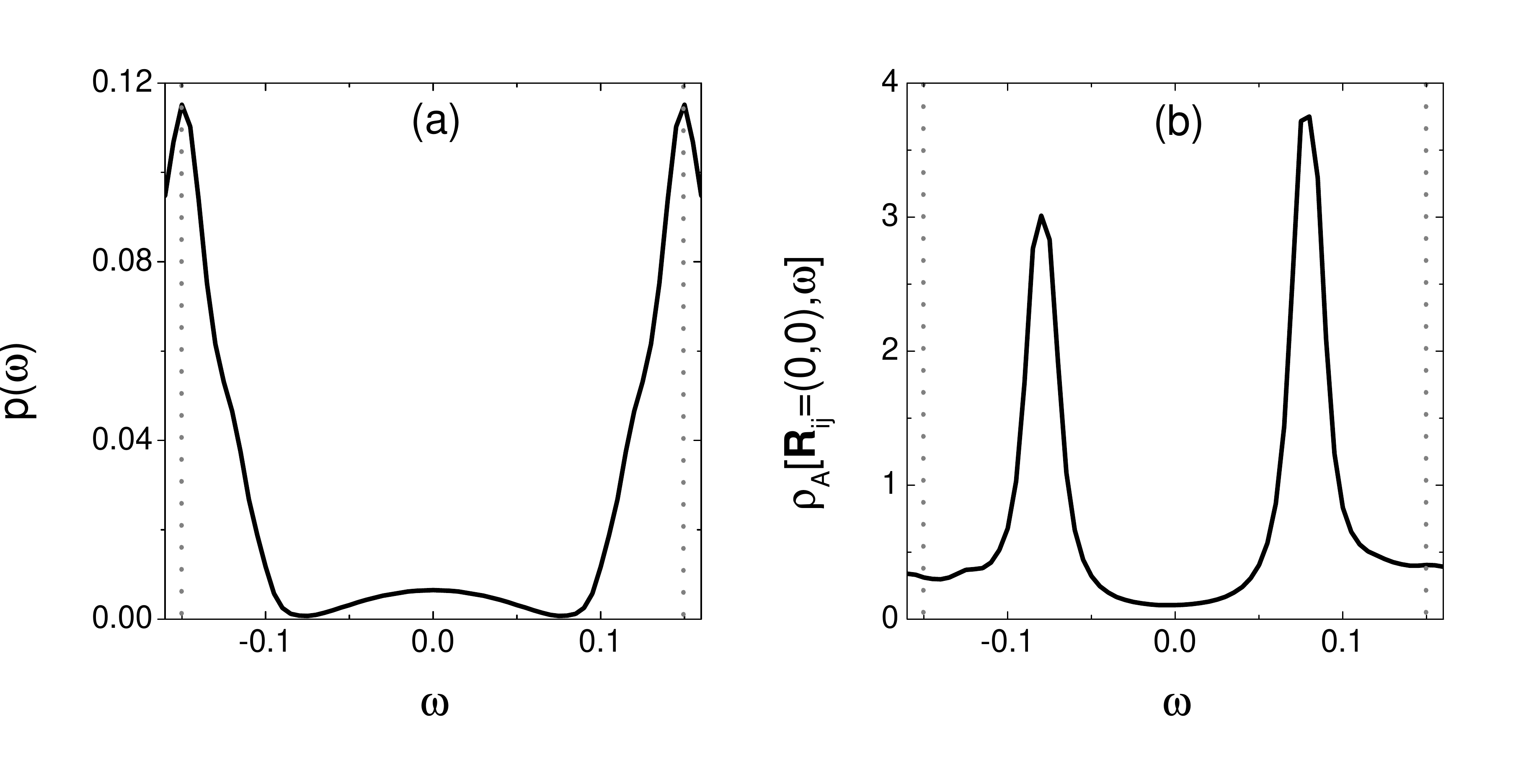}
 \caption{\label{bound1} For the $s^{*}$-wave pairing symmetry. (a) $p(\omega)$ at $V_{s}=5$. (b) The LDOS at the impurity site. The gray dotted lines in both (a) and (b) denote the location of the SC coherence peaks.}
\end{figure}

In fully gapped superconductors, the poles of $T^{r/a}(\omega)$ below the SC gap determine the location of the impurity-induced in-gap bound states \cite{zhujx}, which should show up when $p(\omega)=det[I-\frac{U}{N}\sum_{\mathbf{q}}g_{0}^{r/a}(\mathbf{q},\omega)]=0$. In Fig. \ref{location1}(a), we plot $\omega_{0}$ as a function of $V_{s}$ where $p(\omega_{0})$ is the minimum of $p(\omega)$ when $\omega$ is between the two SC coherence peaks shown in Fig. \ref{gap&dos}(b). We found that $p(\omega_{0})=p(-\omega_{0})$, suggesting that the in-gap bound states, if exist, will always appear in pairs and their locations will be symmetric with respect to $\omega=0$. So in Fig. \ref{location1}, we show only the result at $\omega_{0}\geq0$. From Fig. \ref{location1}(b) we can see that from $V_{s}=3$ to $7$, $p(\omega_{0})\approx0$, so in-gap bound states should show up. In Fig. \ref{bound1} we take $V_{s}=5$ as an example. Indeed, on the impurity site, there are two impurity-induced in-gap states whose locations are exactly the same as $p(\omega)$ reaches its minimum. The intensities of these two states far exceed those of the SC coherence peaks and they are located at about half of the SC gap. Similar behaviors exist for $V_{s}=3$ to $7$, with the locations of these in-gap states being away from the gap edge. Therefore these in-gap states should easily be observed in STM experiments.

\begin{figure}
\includegraphics[width=1\linewidth]{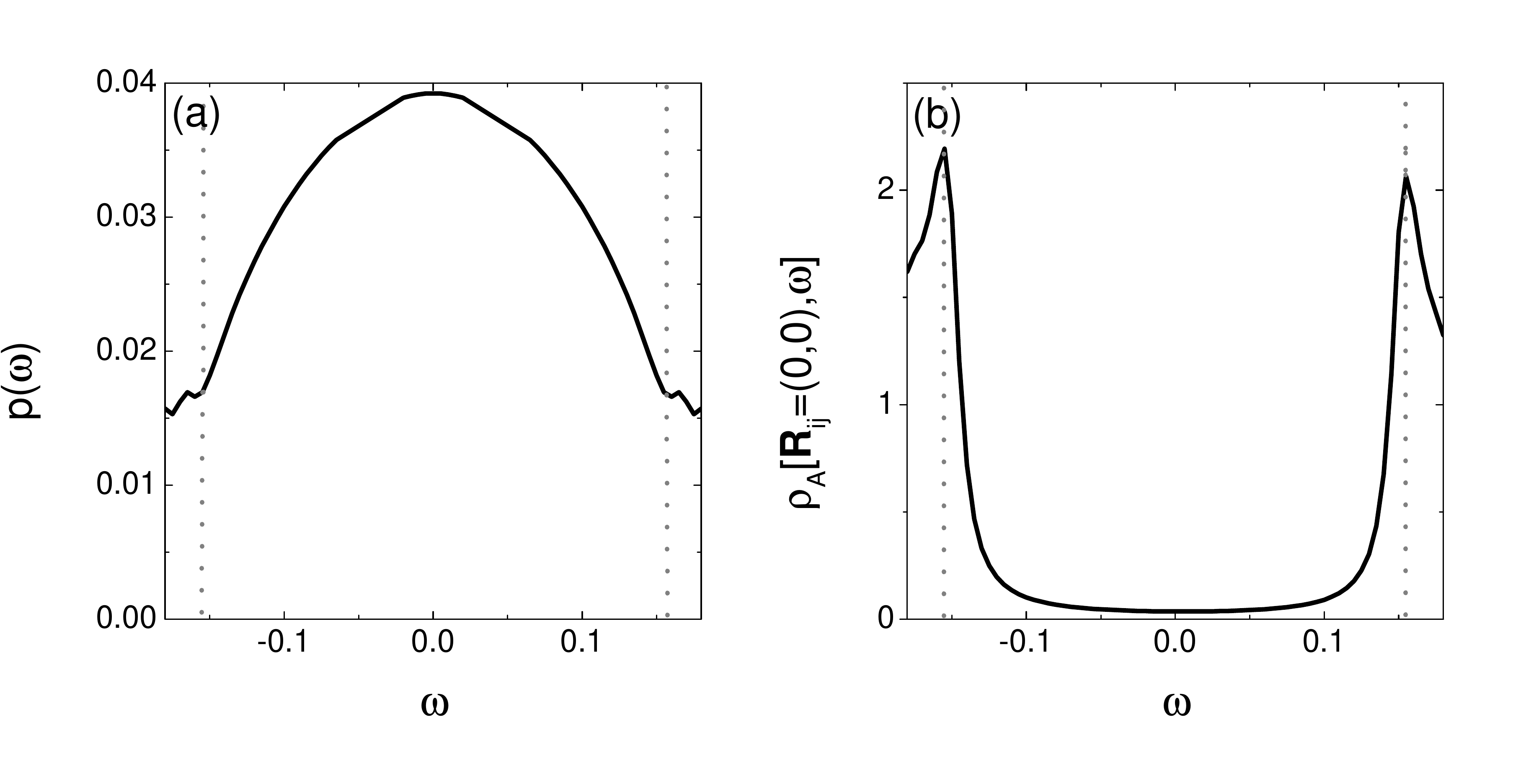}
 \caption{\label{bound2} The same as Fig. \ref{bound1}, but for the conventional $s$-wave pairing symmetry and at $V_{s}=4$.}
\end{figure}

In contrast, if we take $\Delta_{1\mathbf{k}}=0$ and $\Delta_{0}=-0.15$ in Eq. (\ref{dk}), then the pairing symmetry is conventional $s$-wave and the DOS is shown in Fig. \ref{gap&dos}(b), which is very similar to the $s^{*}$ pairing case, except for a higher intensity of the SC coherence peaks. However in this case, after repeating the above calculation we found that $\omega_{0}$ is always located at the gap edges so there are no in-gap bound states and this is shown in Fig. \ref{bound2} where we take $V_{s}=4$ as an example (for $V_{s}=1$ to $9$, the behavior is similar). We can see that although the intensity of the SC coherence peaks is greatly enhanced, there are no in-gap bound states, in sharp contrast to the $s^{*}$ pairing case.

In summary, we construct a tight-binding model suitable for describing the band structure of 1ML FeSe in the absence of the glide mirror symmetry. Then we propose a possible pairing function that can well describe the gap anisotropy observed by ARPES and based on this pairing function we further study the nonmagnetic impurity-induced bound states. The pairing function we proposed has a hidden sign-changing characteristic and clear in-gap bound states can be induced by a nonmagnetic impurity, while in the conventional $s$-wave pairing case, no in-gap bound states exist. Therefore with the help of the STM experiments, the $s^{*}$-wave pairing can be clearly distinguished from the conventional $s$-wave pairing symmetry.

This work is supported by NSFC (Grant No. 11374005) and NSF of Shanghai (Grant No. 13ZR1415400). QHW is supported by NSFC (under grant No.11574134).

\end{document}